# Accurate Determination of the Band-Gap Energy of the Rare-Earth Niobate Series

Alka B. Garg[1,2], David Vie[3], Placida Rodriguez-Hernandez[4], Alfonso Muñoz[4], Alfredo Segura[5], Daniel Errandonea[5,*]

[1]High Pressure and Synchrotron Radiation Physics Division, Bhabha Atomic Research Centre, Mumbai 400085, India

[2]Homi Bhabha National Institute, Anushaktinagar, Mumbai 400094, India

[3]Institut de Ciència dels Materials de la Universitat de València, Apartado de Correos 2085, E-46071 València, Spain

[4]Departamento de Física, Instituto de Materiales y Nanotecnología, MALTA Consolider Team, Universidad de La Laguna, La Laguna, E-38204 Tenerife, Spain

[5]Departamento de Física Aplicada-ICMUV, MALTA Consolider Team, Universidad de Valencia, Edificio de Investigación, Carrer del Dr. Moliner 50, Burjassot, 46100 Valencia, Spain

**Abstract:** In this work, we report diffuse reflectivity measurements in $InNbO_4$, $ScNbO_4$, $YNbO_4$, and eight different rare-earth niobates. From a comparison with the established values of the band gap of $InNbO_4$ and $ScNbO_4$, we have found that the broadly used Tauc plot analysis leads to erroneous estimates of the band-gap energy of niobates. In contrast, accurate results are obtained considering excitonic contributions using the Elliot-Toyozawa model. We have found that $YNbO_4$ and the rare-earth niobates are wide band-gap materials. The band-gap energy is 3.25 eV for $CeNbO_4$, 4.35 eV for $LaNbO_4$, 4.5 eV for $YNbO_4$, and 4.73 – 4.93 eV for $SmNbO_4$, $EuNbO_4$, $GdNbO_4$, $DyNbO_4$, $HoNbO_4$, and $YbNbO_4$. An explanation for the obtained results will be presented. The fact that the band-gap energy is nearly not affected by the rare-earth substitution from $SmNbO_4$ to $YbNbO_4$ and the circumstance that these are the compounds with the largest band gap are a consequence of the fact that the band structure near the Fermi level originates mainly from Nb 4d and O 2p orbitals. We hypothesize that $YNbO_4$, $CeVO_4$, and $LaNbO_4$ have smaller band gaps because of the contribution from rare-earth atom 4d or 5f states to the states near the Fermi level.

TOC IMAGE

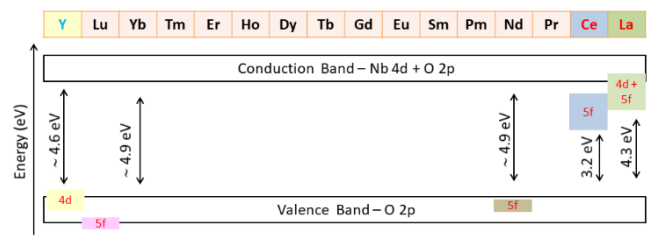

Rare-earth and trivalent metals niobates ($MNbO_4$) have been used for half a century as luminescent materials [1]. These compounds crystallize in the monoclinic fergusonite structure (space group $I2_1/a$ also described with other settings as $I2_1/b$ or $C2/c$) and have exceptional chemical stability, luminescent, and dielectric properties, and ion conductivities [2]. Because of it, they have been proposed for many different multifunctional applications. They include lasers, light emitters, capacitors, optical fibers, medical applications, temperature detectors, bioimaging, photocatalysts for both contaminant degeneration and $H_2$ generation, chemically robust hosts for nuclear materials and wastes, ion conductors for lithium batteries, or solid-oxide fuel cells, among others [2 – 6]. The fact that $MNbO_4$ niobates can be obtained as single crystals [7] and nanocrystals [8] give them a great versatility regarding applications. One of the applications of $MNbO_4$ niobates that has been gaining momentum in the last years is their use for environmentally friendly white-light-emitting diodes (LEDs) [9, 10]. In the last years, these devices have replaced in the market the conventional light sources, including Edison's incandescent lamp and the Hg-discharge-based fluorescent lamp. LEDs have many advantages over the other light sources, including lower power consumption, longer lifetime, improved physical robustness, smaller size, and faster switching [11]. One of the crucial facts to further optimize white LEDs based on orthoniobates, it is the precise determination of the energy of their electronic band gap ($E_g$), which is lacking yet. Several efforts have been devoted to it, but the information in the literature [2, 4, 5, 12 – 16] is scarce and in some cases, contradictory as can be seen in Table 1. Notice that band-gap values from 2.9 to 5.04 eV have been reported for orthoniobates. Such a large variation of $E_{gap}$ is in contradiction with the know-systematic of related orthovanadates, which have been extensively studied [17, 18]. In special, in $YNbO_4$ ($E_{gap}$ = 3.7 – 4.96 eV), $LuNbO_4$ ($E_{gap}$ = 4.2 – 5.04 eV), and $GdNbO_4$ ($E_{gap}$ = 3.48 – 4.89 eV) there are significant discrepancies in the literature. These facts indicate that a systematic study of the band-gap energy of $MNbO_4$ niobates is timely. Here we report diffuse reflectance measurements in $YNbO_4$, $LaNbO_4$, $CeNbO_4$, $SmNbO_4$, $EuNbO_4$, $GdNbO_4$, $DyNbO_4$, $HoNbO_4$, and $YbNbO_4$. A systematic analysis of the results using an Elliot-Toyozawa model [19, 20] has allowed us to accurately determine the band-gap energy of the studied compounds. It allows us also to conclude that the traditional method of determination based on the Tauc plot [21] underestimates the band-gap energy. Our method for the determination of $E_g$ has been validated by performing diffuse reflectance measurements in InNbO$_4$ and ScNbO$_4$, giving an excellent agreement with the established band-gap values, 4.7 and 4.8 eV, respectively [22, 23]. From our study, we conclude that the most studied compounds have a band-gap energy of 4.73 – 4.93 eV, having only CeNbO$_4$ (3.25 eV), LaNbO$_4$ (4.35 eV), and YNbO$_4$ (4.55 eV) narrower band gaps. The observed results will be explained using available band-structure calculations.

Let us start presenting results from InNbO$_4$ and ScNbO$_4$, which have been used to establish the method used to determine $E_g$. Both materials have a monoclinic wolframite-type structure (related to fergusonite) and their band-gap energies have been accurately determined previously [22, 23]. In Fig. 1 we present the absorption spectra $F(R_\infty)$ obtained from the diffuse-reflectance measurement via the Kubelka–Munk transformation [24], which can be considered approximately proportional to the absorption coefficient ($\alpha$) [25]. In both InNbO$_4$ and ScNbO$_4$, the absorbance has a sharp absorption onset above 4 GPa, reaching a maximum at 4.7 and 4.75 eV, respectively (see fig. 1). In the literature, there is an agreement that InNbO$_4$ and ScNbO$_4$ have direct band gaps with $E_{gap}$ = 4.7 and 4.8 eV, respectively. However, if we analyze our results using the traditional Tauc plot analysis [21], determining $E_g$ from a linear least-squares fit to zero in $(h\upsilon F(R_\infty))^2$ versus $h\upsilon$ (as shown in Fig. 1), we found values for $E_{gap}$ which underestimate by 0.4 eV from the established values of $E_{gap}$. We are not surprised by this fact because this method has been recently challenged [26]. The main drawback is that it tends to underestimate the band-gap energy in materials showing sub-band-gap absorption tails related to defects, surface effects, and other phenomena which are reflected in the absorption spectrum as an Urbach tail [26]. The fact that the Tauc analysis neglects the presence of excitons could also lead to the underestimation of $E_{gap}$ [27]. In materials related to MNbO$_4$ niobates, like InVO$_4$, the application of the traditional Tauc plot analysis leads to underestimations of up to 1.8 eV in the value of $E_g$ [28]. Then, it is not surprising that in InNbO$_4$ and ScNbO$_4$, it could lead to an underestimation of 0.4 eV. We will show now that this is because of the excitonic contribution to the fundamental absorption spectrum. In InNbO$_4$ and ScNbO$_4$ the steplike absorption present above 4 eV (see Fig. 1) is typical of direct transitions in which excitonic effects are observed even at room temperature [29]. This can be confirmed by fitting the absorbance obtained by means of the excitonic Elliott-Toyozawa model [19, 20] where:

$$F(R_\infty) \propto \alpha(h\nu)$$

$$\propto \frac{1}{h\nu}\left[\sum_j \frac{2E_B}{j^3} sech\left(\frac{h\nu - E_j^B}{\Gamma}\right)\right.$$

$$\left. + \int_{E_g}^{\infty} sech\left(\frac{h\nu - E}{\Gamma}\right) \frac{1}{1 - \exp(-2\pi\sqrt{\frac{E_g}{E-E_g}})} dE\right]$$

In this expression, the first term within the brackets is the discrete excitonic contribution, and the second term is the continuum contribution. In our fitting we have assumed that only the fundamental exciton state (j = 1) contributes to the peak, being $E_1^B = E_{gap} - E_B$, where $E_B$ is the exciton binding energy. The only fitting parameters are $E_g$, $E_B$, and $\Gamma$, which takes the spectral linewidth into account. This simple model can reproduce the absorbance of the two compounds as it can be seen in Fig. 1. From our fit we have determined $E_{gap}$ = 4.7 for InNbO$_4$ and 4.8 eV for ScNbO$_4$. These values are in excellent agreement with the literature [22, 23]. The obtained values for $E_B$ are 70 and 120 meV, respectively. These values are comparable to the exciton binding energy of wide band-gap oxides [29]. The excellent agreement with the literature and the good quality of fits (using a simple model with only the fundamental exciton) confirm that excitonic contributions are important in the absorbance of MNbO$_4$ niobates. We consequently will use the same model to analyze all the rare-earth niobates we have investigated.

Once the correct method for determining the band-gap energy has been established we will present the results on rare-earth niobates. We will first present SmNbO$_4$, EuNbO$_4$, and GdNbO$_4$. The results of our experiments and the analysis are provided in Fig. 2. The three compounds have a sharp absorption edge typical of a direct band gap. Below that energy, there is a sub-band-gap Urbach tail, typical of MXO$_4$ [30] oxides and week peaks which can be attributed to internal 4f-4f transitions associated to the lanthanide cation [31, 32]. By applying the Elliott-Toyozawa model the fundamental absorption edge can be fitted quite well. The determined band-gap energies are 4.95, 4.73, and 4.93 eV for SmNbO$_4$, EuNbO$_4$, and GdNbO$_4$, respectively. The obtained values for the exciton binding energy (see Fig. 2) are comparable to the values determined for InNbO$_4$ and ScNbO$_4$. The band-gap energy is very close to the energy of the maximum of $F(R_\infty)$ as shown in Fig. 2. In the figure, we also include a fit using the Tauc plot. The values estimated for E$_{gap}$ with this method is always more than 0.4 eV compared to the correct value as can be seen in Table 1. In this table, we also compare our results with previous studies. For SmNbO$_4$, this is the first determination of E$_{gap}$, the value obtained is consistent with photoluminescence measurements, that constrain the band-gap energy to 4.7-5.0 eV [33]. In the case of GdNbO$_4$, our result (4.93 eV) is in excellent agreement with the results of Feng et al. (4.89 eV) [12]. This indicates that the band gap of 3.48 eV reported by Hirano et al. is an underestimation. The same can be stated by the band gap of 3.45 eV determined by the same authors for EuNbO$_4$ [12], which is 1.3 eV smaller than the value of E$_{gap}$ determined in the present work.

In Fig. 3 we present the results obtained from DyNbO4, HoNbO4, and YbNbO4. Again, the three compounds have a very large band gap E$_{gap}$ ~ 4.9 eV (see Fig. 3 and Table 1). In Fig. 3 it can also be seen that the band-gap energy is very close to the energy of the maximum of $F(R_\infty)$. The figure also shows how E$_{gap}$ is underestimated by ~ 0.4 eV when the Tauc plot analysis is applied. Notice that E$_{gap}$ in the three compounds is very similar to E$_{gap}$ in SmNbO$_4$, EuNbO$_4$, and GdNbO$_4$. As we will explain towards the end of the manuscript, this is a consequence of the fact that the electronic states at the bottom of the conduction band and the top of the valence band are dominated by O 2p states and Nb 4d states. Thus, as a first approximation, the band gap is determined by the configuration of the NbO$_4$ tetrahedron, which does not change significantly from one compound to the other. The three of them also have the typical Urbach tail plus a contribution from absorptions from the 4f levels of the rare-earth atoms. In addition, HoNbO4 and DyNbO$_4$ have in $F(R_\infty)$ sub-band-gap sharp absorptions caused by internal 4f-4f transitions of Dy and Ho [32]. For DyNbO$_4$ and HoNbO$_4$, this is the first time that the band-energy is reported. In YbNbO$_4$, our result (E$_{gap}$ = 4.9 eV) showes that E$_{gap}$ has been largely underestimated in a previous report [16].

In Fig. 4, we present the results obtained from YNbO$_4$, LaNbO$_4$, and CeNbO$_4$. The figure shows that the Tauc analysis always gives underestimated values of E$_{gap}$.

According to the Elliot-Toyozawa model, YNbO$_4$ has a band-gap energy of 4.56 eV (see Fig. 4 and Table 1). This value is 10% smaller than in the previously discussed niobates, and it is comparable to the maximum value of E$_{gap}$ reported in previous studies [2, 4, 5, 14]. For LaNbO$_4$, we have determined E$_{gap}$ 4.35 eV. The previously determined value was 10% smaller [13]. For CeNbO$_4$, this is the first time that E$_{gap}$ is determined. The obtained value, 3.25 eV, indicates that this compound is the orthoniobate with the smallest band gap. As in the other studied compounds, in the three compounds presented in Fig. 4, the band-gap energy is very close to the energy of the maximum of $F(R_\infty)$.

It is interesting to stress that the band-gap energy of rare-earth niobates follows a very similar trend as observed for rare-earth vanadates [17, 34, 35]. In the vanadates, except for LaVO$_4$ and CeVO$_4$, all the members of the family have a band gap close to 3.8 eV. LaVO$_4$ has a band gap of 3.5 eV and CeVO$_4$ a band gap of 3.2 eV. The 3.8 eV band gap is a consequence of the fact that states near the Fermi level are dominated by O 2 p and V 3 d states. In the other compounds, the band gap is reduced due to the contribution of 5d and 4f electrons from La and Ce. A similar picture provides a qualitative explanation to the results reported here. The band structure and electronic density of states (DOS) of most niobates are reported in the literature [2, 13, 36, 37, 38]. They are qualitatively similar to the band structure and DOS of vanadates [34], having the band structures a similar topology and the states near the Fermi level a similar composition. According to band-structure calculations, the studied niobates have an indirect gap, but the difference with the direct band gap is smaller than 0.1 eV [2, 13, 36, 37, 38]. Then the absorption edge will be dominated by the direct band-gap and $F(R_\infty)$ can be analyzed assuming a direct band gap as it has been done in this work.

To provide a more quantitative discussion, we will focus on LaNbO$_4$, CeNbO$_4$, EuNbO$_4$, and YbNbO$_4$ [38], which are representative of the family of fergusonite-type niobates. In the four compounds, the minimum of the conduction band (CB) and the maximum of valence band (VB) are located at V and $\Gamma$ point of the Brillouin zone. However, there is a $\Gamma$–$\Gamma$ direct band gap very close in energy. According to the electronic DOS, in EuNbO$_4$ and YbNbO$_4$ the valence-band maximum is dominated by O 2p states. This behavior is typical of most wide-band-gap oxides. On the other hand, the conduction-band minimum is predominantly dominated by Nb 4d states with a small contribution of O 2p states. This molecular orbital composition is analogous to what happens in vanadates with V 3d and O 2p states [34] and in other ternary oxides like tungstates (molybdates) with W (Mo) 5d (4d) and O 2p states [39]. Consequently, the configuration of the $NbO_4$ tetrahedron, the coordination polyhedron of Nb in ferguson­ite, will have a determinant role in determining the band-gap energy. Interestingly, this unit is little modified when going from one compound to the other, changing the average Nb-O bond distance less than 1% along the series of rare-earth niobates. Then, it is not surprising that most members of this family have a very similar $E_{gap}$. It is also not surprising that these compounds have a band-gap energy comparable to that of $Nb_2O_5$ ($E_{gap}$ = 5.1 eV) in which states at the bottom of the CB and top the VB are also dominated by the hybridization of O 2p and Nb 4d states, and O 2p states, respectively [40]. The reason to the band-gap energy closing of $LaNbO_4$ and $CeNbO_4$ comes from the fact that these are the only two compounds where the lanthanide orbitals contribute to the conduction band. Exactly as happens in $LaVO_4$ and $CeVO_4$, the vanadates with the smallest band-gap energy [17].

Summing up, in this letter we reported diffuse reflectance measurements for rare-earth niobates and proposed a method for accurately determining their band-gap energy. We have shown that in most previous studies the band-gap energy has been underestimated. The band-gap energy of the studied compounds is 3.25 eV for $CeNbO_4$, 4.35 eV for $LaNbO_4$, 4.5 eV for $YNbO_4$, and 4.73 – 4.93 eV for $SmNbO_4$, $EuNbO_4$, $GdNbO_4$, $DyNbO_4$, $HoNbO_4$, and $YbNbO_4$. We also provide an explanation to the experimental results based on density functional calculations. The reported information is crucial for technological applications of rare-earth niobates.

**METHODS**

**Experimental details:** Polycrystalline rare-earth niobates were synthesized by the commonly used solid-state reaction technique. Starting metal oxides, $R_2O_3$ and $Nb_2O_5$ (purity > 99.9%) were pre-dried to remove any moisture or organic impurities. These dried binary oxides were weighed in stoichiometric (1:1) ratio, followed by hand mixing in pestle and mortar, cold pressing into cylinders of 12.5 mm in diameter and 5 mm in height, and fired at 1200 °C for 24 hrs. in a box type programmable resistive furnace. These pellets were further sintered at 1300 °C for 48 hrs. Single phase formation of the compound was confirmed by angle-dispersive powder x-ray diffraction. We confirmed that all synthesized rare-earth niobates crystallize in the fergusonite structure. For measurements in $ScNbO_4$ and $InNbO_4$, we used the same samples used in previous studies [21, 22]. For the diffuse reflectance measurements, powder samples were manually grounded in an agata mortar, placed in the sample holder, and supported with a quartz window. Measurements were carried out on a Shimazdu UV-Vis 2501PC spectrophotometer equipped with an integrating sphere for diffuse reflectance measurements. $BaSO_4$ was used as reference material and for background measurement. The spectral range covered by measurements was 220-900 nm, with a resolution of 1 nm and a Spectral bandwidth of 5 nm. The spectra recorded in %R were transformed using the Kubelka-Munk function [24].

## AUTHORS INFORMATION

**Corresponding Author**

*E-mail: daniel.errandonea@uv.es

ORCID: 0000-0003-0189-4221

**Author Contributions**

A.B.G. synthesizes the samples. D.V. performed the diffuse reflectance measurements. D.E. designed the study and carried out the data analysis. All authors contribute to the discussion, writing, and proofreading and have given approval to the final version of the manuscript.

**Funding Sources**

Spanish Mineco (MINECO/AEI/0.13039/501100003329) Project No. RED2018-102612-T, Spanish MCIN (MCIN/AEI/10.13039/501100011033) Projects No. PID2019-106383GB-41/43, Generalitat Valencina Project No. PROMETEO CIPROM/2021/075 (GREENMAT), and Advanced Materials Programme supported by MCIN with funding from European Union NextGenerationEU (PRTR-C17.I1) Project MFA/2022/007.

**Notes**

The authors declare no competing financial interest.

**ACKNOWLEDGMENTS**

This study was supported by the MALTA Consolider Team network (Project No. RED2018-102612-T), financed by MINECO/AEI/0.13039/501100003329, the I+D+i Project No. PID2019-106383GB-41/43 financed by MCIN/AEI/10.13039/501100011033, as well as by the Project No. PROMETEO CIPROM/2021/075 (GREENMAT) financed by Generalitat Valenciana. This study forms part of the Advanced Materials Programme and was supported by MCIN with funding from European Union NextGenerationEU (PRTR-C17.I1) and by Generalitat Valenciana under grant MFA/2022/007. The authors also thank Prof. Ana Costero from Dept. de Química Orgánica at Univ. de Valencia for letting us use the Shimazdu UV-Vis 2501PC spectrophotometer for diffuse reflectance measurements.

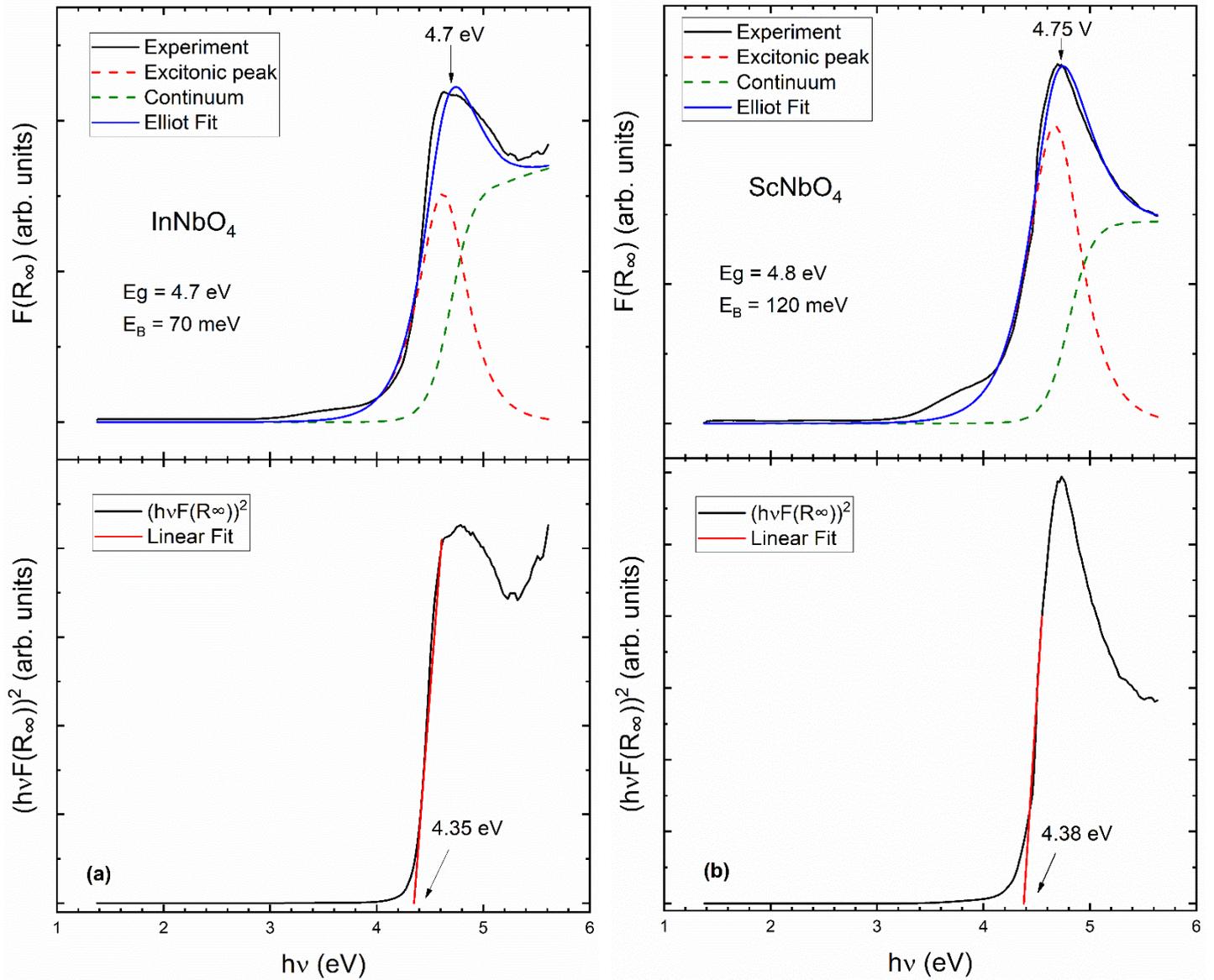

**Figure 1:** (a) In the top we plot the $F(R_\infty)$ spectrum of InNbO$_4$ together with the fit used to determine the band-gap energy. The black line is the experiment, the blue line is the fit, the green dashed line is the continuum contribution, and the red dashed lie is the excitonic contribution. In the bottom we show the Tauc plot traditionally used to determine E$_{gap}$. (b) Same but for ScNbO$_4$.

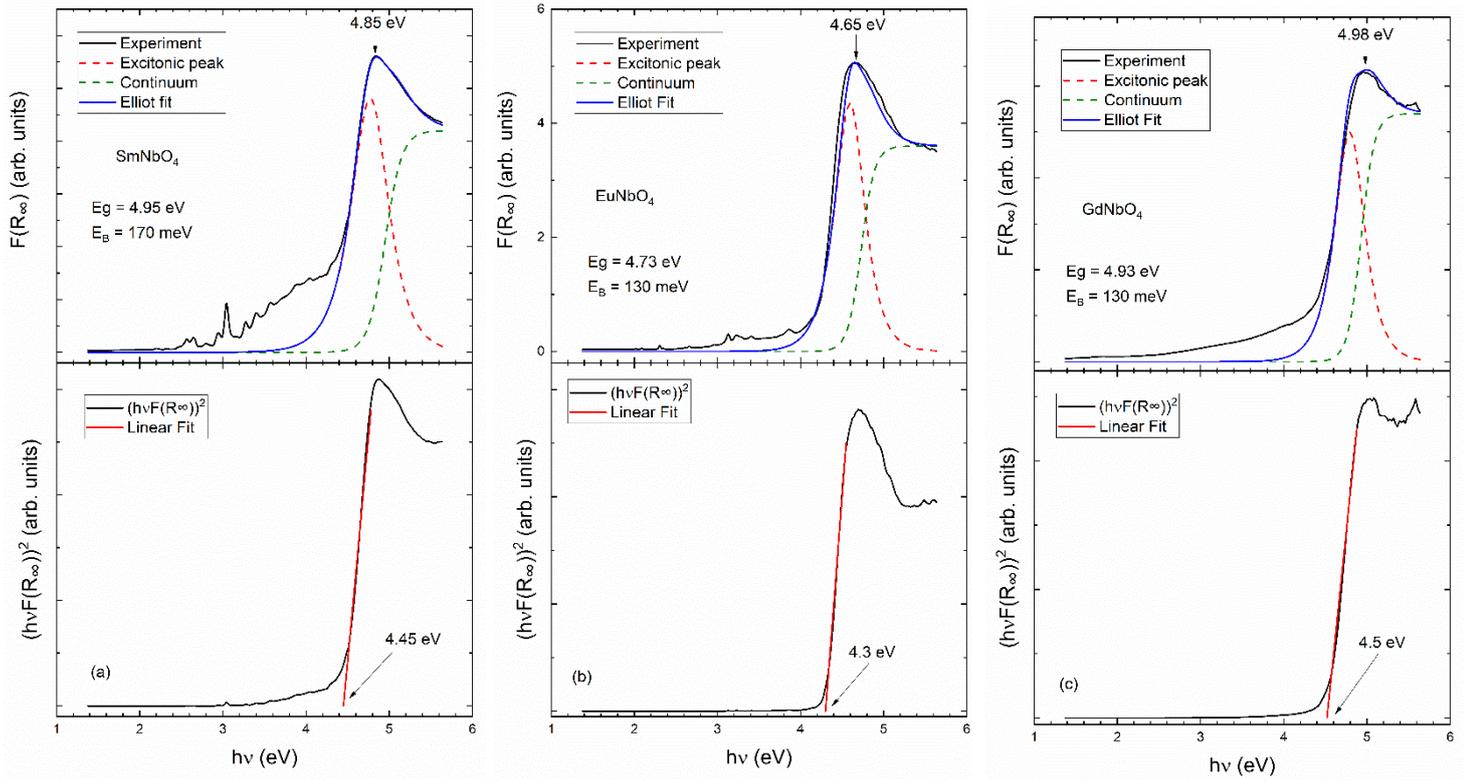

**Figure 2:** (a) In the top we plot the $F(R_\infty)$ spectrum of SmNbO₄ together with the fit used to determine the band-gap energy. The black line is the experiment, the blue line is the fit, the green dashed line is the continuum contribution, and the red dashed lie is the excitonic contribution. In the bottom we show the Tauc plot traditionally used to determine E$_{gap}$. (b) Same but for EuNbO₄. (c) Same but for GdNbO₄.

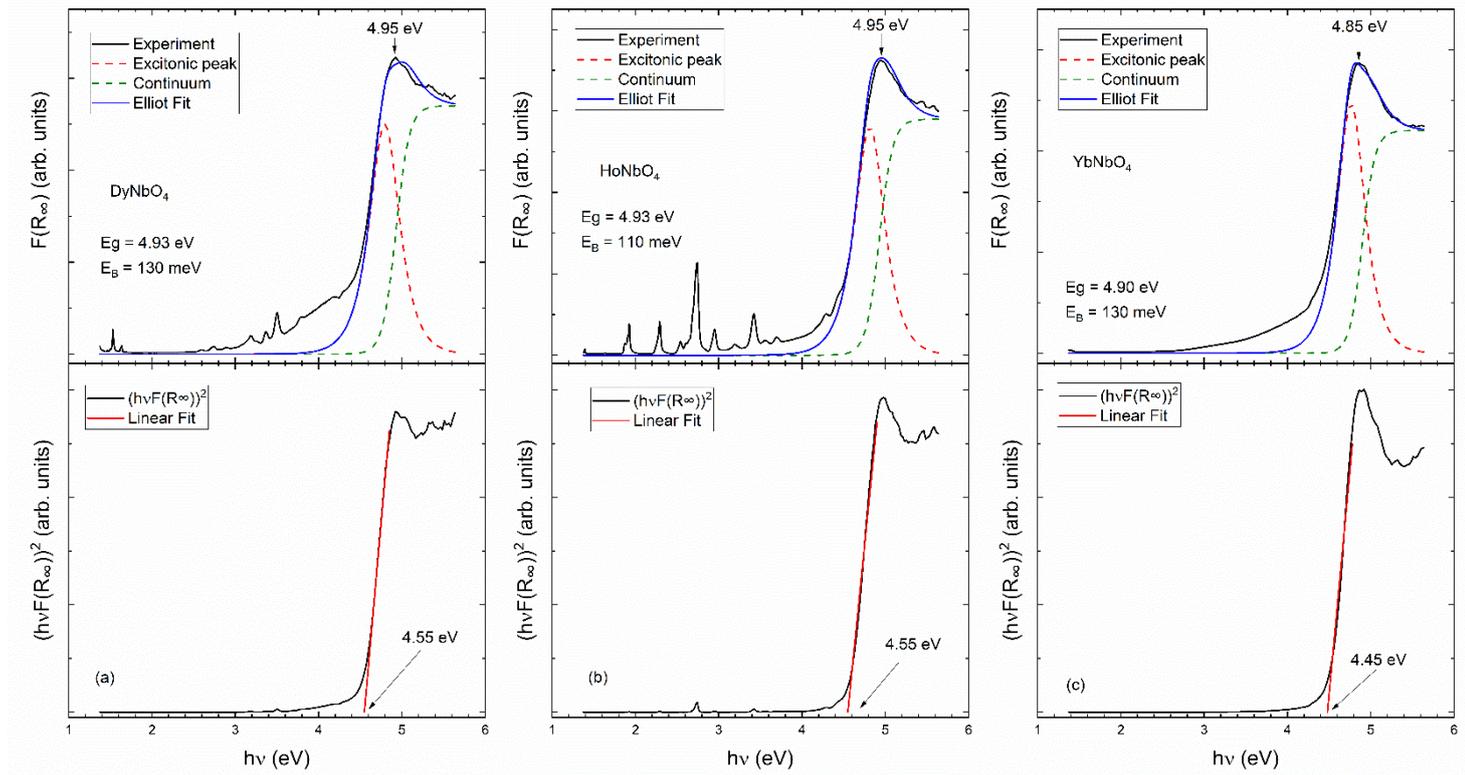

**Figure 3:** (a) In the top we plot the $F(R_\infty)$ spectrum of DyNbO$_4$ together with the fit used to determine the band-gap energy. The black line is the experiment, the blue line is the fit, the green dashed line is the continuum contribution, and the red dashed lie is the excitonic contribution. In the bottom we show the Tauc plot traditionally used to determine E$_{gap}$. (b) Same but for HoNbO$_4$. (c) Same but for YbNbO$_4$.

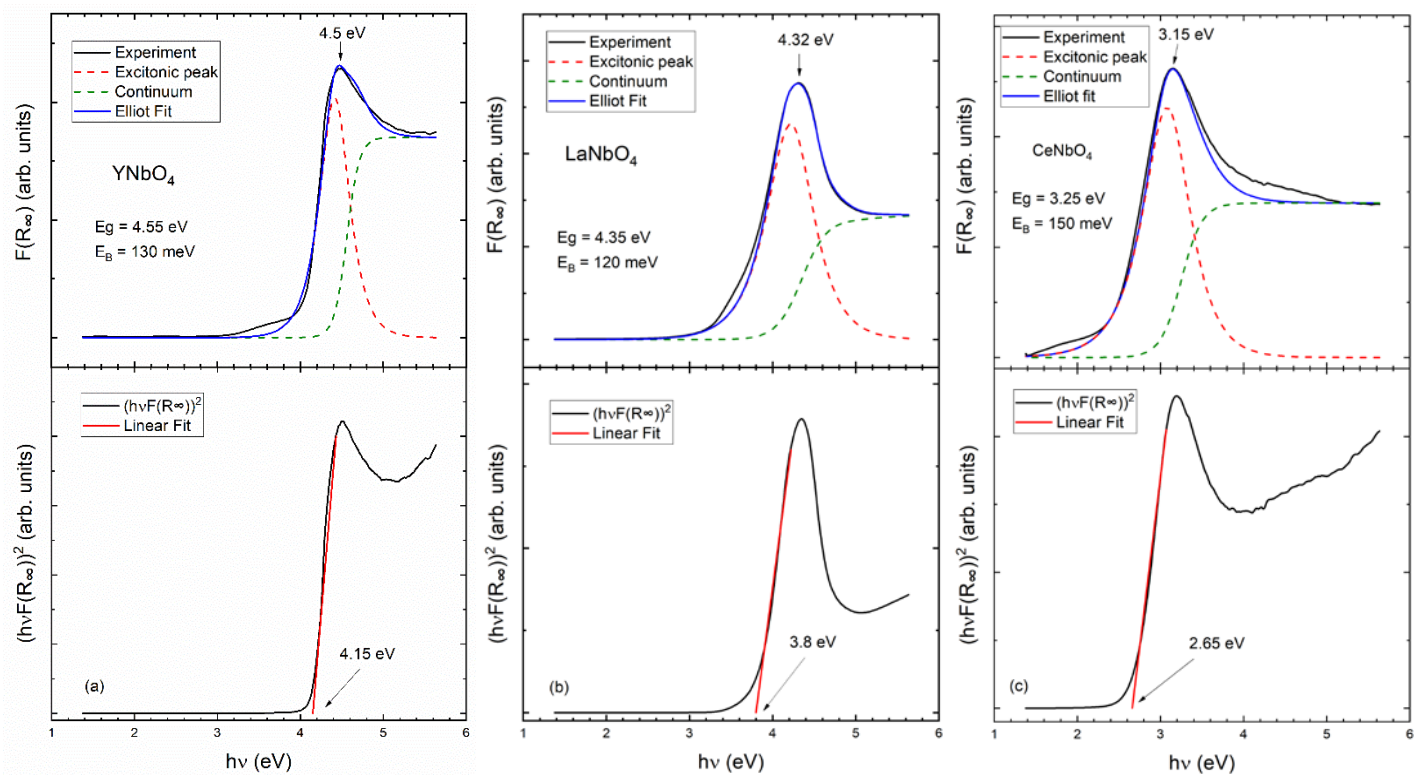

**Figure 4:** (a) In the top we plot the $F(R_\infty)$ spectrum of YNbO$_4$ together with the fit used to determine the band-gap energy. The black line is the experiment, the blue line is the fit, the green dashed line is the continuum contribution, and the red dashed lie is the excitonic contribution. In the bottom we show the Tauc plot traditionally used to determine E$_{gap}$. (b) Same but for LaNbO$_4$. (c) Same but for CeNbO$_4$.

**Table 1:** Band-gap energy of orthoniobates given in eV. Results obtained using the Elliot model and the Tauc plot are compared with the literature.

| Compound | Elliot Fit | Tauc Plot | Literature | Reference |
|---|---|---|---|---|
| $InNbO_4$ | 4.7 | 4.35 | 4.7 | [22] |
| $ScNbO_4$ | 4.8 | 4.38 | 4.8 | [23] |
| $YNbO_4$ | 4.55 | 4.15 | 3.7 – 4.96 | [2, 4, 5, 14] |
| $LaNbO_4$ | 4.35 | 3.80 | 4.0 | [13] |
| $CeNbO_4$ | 3.25 | 2.65 | | |
| $PrNbO_4$ | | | 4.8 | [15] |
| $NdNbO_4$ | | | | |
| $SmNbO_4$ | 4.95 | 4.45 | 4.7 - 5.0 | [33] |
| $EuNbO_4$ | 4.73 | 4.30 | 3.45 | [12] |
| $GdNbO_4$ | 4.93 | 4.50 | 3.48 – 4.89 | [4, 12] |
| $TbNbO_4$ | | | 2.9 | [5] |
| $DyNbO_4$ | 4.93 | 4.55 | | |
| $HoNbO_4$ | 4.93 | 4.55 | | |
| $ErNbO_4$ | | | 3.5 | [16] |
| $TmNbO_4$ | | | | |
| $YbNbO_4$ | 4.90 | 4.45 | 3.46 | [16] |
| $LuNbO_4$ | | | 4.2 – 5.04 | [2, 4] |